\documentclass[12pt]{article}

\usepackage[a4paper,left=30mm,right=20mm,top=20mm,bottom=20mm]{geometry}
\usepackage{graphics}
\usepackage{epsfig}
\usepackage{cite}

\title{Drell-Yan lepton pair production at LHC and \\TMD quark densities of the proton}

\author{S.P.~Baranov$^1$, A.V.~Lipatov$^{2,\,3}$, N.P.~Zotov$^2$}

\begin{document}

\maketitle

\begin{center}

{\it $^1$P.N.~Lebedev Institute of Physics, 119991 Moscow, Russia}\\
{\it $^2$Skobeltsyn Institute of Nuclear Physics, Lomonosov Moscow State University, 119991 Moscow, Russia}\\
{\it $^3$Joint Institute for Nuclear Research, 141980 Dubna, Moscow Region, Russia}

\end{center}

\vspace{0.5cm}

\begin{center}

{\bf Abstract }

\end{center}

We use the TMD quark densities of the proton to
investigate unpolarized Drell-Yan lepton pair production in 
proton-proton collisions at the LHC energies.
We investigate the case where the gluon-to-quark splitting occurs at the last evolution step
and calculate the TMD sea quark density as a convolution of the CCFM-evolved gluon 
distribution and the TMD gluon-to-quark splitting function
which contains all single logarithmic small-$x$ corrections
to the sea quark evolution for any order of perturbation theory.
Based on the ${\cal O}(\alpha)$ production 
amplitude $q^* + \bar q^* \to Z/\gamma^* \to l^+ + l^-$ which
calculated according to the reggeized quark approach,
we analyze the distributions on the dilepton invariant mass,
transverse momentum and rapidity as well as 
the specific angular correlations between the produced leptons 
as measured by the CMS, ATLAS and LHCb collaborations.
We argue that these measurements impose stringent 
constraints on the TMD quark distributions of the proton.

\vspace{1.0cm}

\noindent
PACS number(s): 12.38.-t, 12.15.Ji

\newpage

\section{Introduction} \indent

Usually, the scale-dependent quark and gluon densities are  
calculated as a function of Bjorken variable $x$
and hard scale $\mu^2$ within the framework of the DGLAP 
evolution equations\cite{1} based on the standard 
collinear QCD factorization. 
However, for a wide class of semi-inclusive processes probing 
the small-$x$ and forward physics at the LHC it is more appropriate
to use the parton distributions unintegrated over the partonic transverse momentum $k_T$ or, 
transverse momentum dependent (TMD) parton distributions\cite{2}.
The latter are subject of intense studies, and various approaches to 
investigate these quantities have been proposed\cite{3,4,5,6}.
At asymptotically large energies (or very small $x$) the theoretically correct 
description is based on the BFKL evolution equation\cite{7} 
where the leading $\ln(1/x)$ contributions are 
taken into account to all orders. 
Another approach, which is valid for both small and large $x$, is given by the CCFM  
gluon evolution equation\cite{8}. In the limit of asymptotic high energies, it is
equivalent to BFKL, but also similar to the DGLAP evolution for large $x $.
Two basic TMD gluon densities are used in the
small-$x$ formalism: the so-called Weizsaker-Williams and
the dipole ones\cite{9,10,11}.

Most of previous phenomenological applications of the TMD parton distributions 
in the framework of high energy QCD factorization 
(or $k_T$-factorization approach)\cite{12,13} take
 only gluon and valence quark contributions 
 into account (see, for example,\cite{14,15,16,17,18}). 
Such approaches are reasonable approximations, based on the dominance of spin-$1$ exchange processes 
at high energies, tfor the production processes coupled to the gluons
(such as, for example, production of heavy flavours or scalar particles). However, 
to correctly treat the final states associated with the
quark-initiated processes it is necessary to
go beyond this simple approximation and take into account 
subleading effects. First attempts to address this issue
and evaluate the TMD sea quark density 
have been performed in\cite{19,20,21,22}, where 
 the relevant TMD gluon distribution has been derived  via splitting probabilities to lowest order 
of perturbative theory, neglecting any transverse momentum dependence
in the gluon-to-quark branching.
In\cite{23} the TMD kinematic corrections have been included,
while the splitting kernels are still taken at lowest order.
Recently, the TMD sea quark densities
have been calculated\cite{24} incorporating the effects of the
TMD gluon-to-quark splitting 
function\cite{25} which contains all single logarithmic small-$x$ corrections
to sea quark evolution for any order of perturbation theory, and
the kinematical effects from initial state parton transverse momentum
on the forward $Z$ boson spectrum have been studied.
The proposed formulation\cite{24, 25,26} has been implemented in a 
Monte Carlo event generator \textsc{cascade}\cite{27}.
In the present paper we apply the TMD sea quark densities\cite{24} 
to investigate the Drell-Yan lepton pair
production at the LHC. This process, where 
quark-antiquark annihilation form a intermediate virtual photon $\gamma^*$ or $Z$ boson 
decaying to lepton pairs, offers high sensitivity to the sea quark
evolution of a proton.

The production of Drell-Yan lepton pairs at the LHC
is subject of intense studies from both
theoretical and experimental points of view\cite{28,29,30,31,32,33,34,35}. 
It provides a major source of background to a number of processes, 
such as Higgs, $t\bar t$-pair, di-boson or $W^\prime$ and $Z^\prime$ bosons 
production and other processes beyond the Standard Model. It is an 
important reference process for measurements of
electroweak boson properties,
and it is used for monitoring of the collider luminosity and
calibration of detectors. A first study of Drell-Yan process
at the TMD level has been performed in\cite{36}, where  diagrams with virtual photon
exchange only are considered and  mostly on the rather low energies 
covered by the RHIC and UA1
experiments are covered.
A more recent investigation\cite{37} was based on the ${\cal O}(\alpha)$ and 
${\cal O}(\alpha \alpha_s)$ production amplitudes of $q + \bar q \to Z/\gamma^* \to l^+ + l^-$ 
and $q + g^* \to Z/\gamma^* + q \to l^+ + l^- + q$ subprocesses where only the initial
gluon transverse momentum has been taken into account.
This process has been investigated also in the framework of  the
soft-collinear effective theory\cite{38,39}, and general investigation  of high energy 
resummation for Drell-Yan lepton pair production
has been done in\cite{40}.  
In the present paper we concentrate on the off-shell (or transverse momentum dependent) 
quark-antiquark annihilation $q^* + \bar q^* \to Z/\gamma^* \to l^+ + l^-$ 
and calculate the corresponding production amplitude according to 
the reggeized quark approach\cite{41,42}, which is based on the effective
action formalism\cite{43}, currently explored at next-to-leading order\cite{44}. 
It was shown that the use of
effective vertices\cite{41,42} ensures the exact gauge invariance of 
calculated amplitude 
despite the off-shell initial quarks.
We apply the TMD valence and sea quark distributions\cite{15,24} to calculate the Drell-Yan production 
cross sections at  LHC energies.
For comparison, we also use the TMD quark densities 
obtained in the  Kimber-Martin-Ryskin (KMR) scheme\cite{20,21}.
We analyze the dilepton transverse momentum and rapidity distributions
as well as the specific angular correlations between the produced leptons
and compare our predictions with recent data taken by the CMS\cite{28,29,30}, ATLAS\cite{31,32,33,34} 
and LHCb\cite{35} collaborations.
Note that we present a first phenomenological application of the formalism developed in \cite{24,25} 
to the analysis of experimental data.

The outline of our paper is following. In Section~2 we recall shortly
the basic formulas of $k_T$-factorization QCD approach. The TMD quark densities 
are discussed in Section~3. In Section~4 we present numerical 
results of our calculations. Section~4 contains our conclusions.

\section{Theoretical framework} \indent

Our consideration is based on the $\cal{O}(\alpha)$ subprocess of off-shell 
quark-antiquark annihilation into a virtual photon or $Z$ boson
which decays to lepton pair:
\begin{equation}
  q^*(q_1) + \bar q^*(q_2) \to Z/\gamma^* \to l^+(p_1) + l^-(p_2),
\end{equation}

\noindent 
where the four-momenta of all corresponding particles are given in the parentheses.
Note that ${\cal{O}}(\alpha \alpha_s)$ contributions from $q^* + g^* \to Z/\gamma^* + q \to l^+ + l^- + q$ and 
$q^* + \bar q^* \to Z/\gamma^* + g \to l^+ + l^- + g$ subprocesses 
are effectively taken into account in our consideration 
due to the initial state gluon radiation. 
This is in  contrast with  collinear QCD factorization 
where all these contributions have to be taken into account 
separately\footnote{See, for example, reviews\cite{2} for more information.}.

Within the reggeized quark formalism\cite{41,42}, the off-shell 
amplitude of $q^* + \bar q^* \to Z/\gamma^* \to l^+ + l^-$ subprocess can be written as
\begin{equation}
  {\cal M}_{\gamma} = e_q e^2 \, \bar v_{s_1}(q_2) \Gamma^{\mu}_{\gamma}(q_1,q_2) u_{s_2}(q_1) {g^{\mu \nu} \over \hat s} \bar u_{r_1}(p_1) \gamma^{\nu} v_{r_2}(p_2),
\end{equation}
\begin{equation}
  \displaystyle {\cal M}_{Z} = {e^2 \over \sin 2\theta_W} \bar v_{s_1}(q_2) \Gamma^{\mu}_{Z}(q_1,q_2) u_{s_2}(q_1) \left[g^{\mu \nu} - { (q_1 + q_2)^\mu (q_1 + q_2)^\nu \over m_Z^2 }\right] \times \atop { 
    \displaystyle \times {1 \over \hat s - m_Z^2 - im_Z\Gamma_Z} \bar u_{r_1}(p_1) \gamma^{\nu} (C_V^l - C_A^l\gamma^5) v_{r_2}(p_2) },
\end{equation}

\noindent 
where $e$ and $e_q$ are the electron and quark (fractional) electric charges, $\hat s = (q_1 + q_2)^2$,
$m_Z$ and $\Gamma_Z$ are the mass and full decay width of $Z$ boson, $\theta_W$ is the 
Weinberg mixing angle, $C_V^l$ and $C_A^l$ are the vector and axial lepton coupling constants,
and the transverse momenta of initial quarks are ${\mathbf q}_{1T}^2 \neq 0$ and 
${\mathbf q}_{2T}^2 \neq 0$.
We take the propagator of the intermediate $Z$ boson in the Breit-Wigner form to avoid any 
artificial singularities in the numerical calculations.
The effective vertex $\Gamma^{\mu}_{\gamma}(q_1,q_2)$ which describes 
the effective coupling of off-shell (reggeized) quark and antiquark to the photon reads\cite{41,42}
\begin{equation}
  \Gamma^{\mu}_{\gamma}(q_1,q_2) = \gamma^\mu - \hat q_1 {l_1^\mu \over q_2 \cdot l_1 } - \hat q_2 {l_2^\mu \over q_1 \cdot l_2 },
\end{equation}

\noindent 
where $l_1$ and $l_2$ are the four-momenta of colliding protons. 
The coupling of the off-shell quark and antiquark to the $Z$ boson is constructed
in a similar way:
\begin{equation}
  \Gamma^{\mu}_{Z}(q_1,q_2) = \Gamma^{\mu}_{\gamma}(q_1,q_2) (C_V^q - C_A^q\gamma^5),
\end{equation}

\noindent 
where $C_V^q$ and $C_A^q$ are the corresponding vector and axial coupling constants.
The effective vertexes $\Gamma^{\mu}_{\gamma}(q_1,q_2)$ and $\Gamma^{\mu}_{Z}(q_1,q_2)$ satisfy 
the Ward identities $\Gamma^{\mu}_{\gamma}(q_1,q_2) (q_1 + q_2)_\mu = 0$ and
$\Gamma^{\mu}_{Z}(q_1,q_2) (q_1 + q_2)_\mu = 0$.
It is
obvious that the amplitudes~(4) and~(5) are gauge invariant despite the 
off-shell initial quarks.  In all other respects the evaluation 
follows the standard QCD Feynman rules. The further calculation (including 
$\gamma^*$ --- $Z$ interference) is straightforward and was done using the algebraic manipulation 
system \textsc{form}\cite{45}. We do not list here explicitly the lengthy  
expressions. In the on-shell limit,
with ${\mathbf q}_{1T}^2 \to 0$ and ${\mathbf q}_{2T}^2 \to 0$,
we  recover the well-known textbook formulas. 

To calculate the total and differential cross sections 
one has to convolute the evaluated off-shell amplitude squared $|\bar {\cal M}|^2$ with the 
TMD quark densities of the proton. Our master formula reads:
\begin{equation}
  \sigma=\sum_q\int\frac{|\bar {\mathcal M}|^2}{16\pi\, (x_1 x_2 s)^2} f_{q}(x_1,\mathbf q_{1T}^2,\mu^2) f_{q}(x_2,\mathbf q_{2T}^2,\mu^2) d\mathbf p_{1T}^2 d\mathbf q_{1T}^2 d\mathbf q_{2T}^2 dy_1 dy_2 \frac{d\phi_1}{2\pi}\frac{d\phi_2}{2\pi},
\end{equation}

\noindent 
where $s$ is the total energy, $y_1$ and $y_2$ are the center-of-mass rapidities of the produced leptons, $\phi_1$ and 
$\phi_2$ are the azimuthal angles of the initial quarks having the fractions $x_1$ and $x_2$
of the longitudinal momenta of the colliding protons. Finally, from the conservation laws one can 
easily obtain the following relations:
\begin{equation}
  \mathbf q_{1T} + \mathbf q_{2T} = \mathbf p_{1T} + \mathbf p_{2T},
\end{equation}
\begin{equation}
  x_1 \sqrt s = m_{1T} e^{y_1} + m_{2T} e^{y_2},
\end{equation}
\begin{equation}
  x_2 \sqrt s = m_{1T} e^{ - y_1} + m_{2T} e^{ - y_2},
\end{equation}

\noindent 
where $\mathbf p_{1T}$ and $\mathbf p_{2T}$ are the transverse momenta of produced leptons, and
$m_{1T}$ and $m_{2T}$ are their transverse masses.

\section{TMD quark densities} \indent

In the present paper we concentrate on the CCFM
approach to calculate the TMD parton densities of the proton.
As it was already mentioned above, the CCFM parton shower,
based on the principle of color coherence, describes only
the emission of gluons, while real quark emissions are left aside.
It implies that the CCFM equation describes only the distinct evolution of 
TMD gluon and valence quarks, while the non-diagonal transitions between quarks and gluons are absent.
The TMD gluon\cite{46} and valence quark\cite{15} 
distributions $f_g(x,{\mathbf k}_T^2,\mu^2)$ and 
$f_q^{(v)}(x,{\mathbf q}_T^2,\mu^2)$ have been obtained
from the numerical solutions of the CCFM equation. 
Here ${\mathbf k}_T$ and ${\mathbf q}_T$ are the 
gluon and quark transverse momenta, respectively.
In the approximation where the sea quarks occur in the last gluon-to-quark splitting,
the TMD sea quark density at the next-to-leading logarithmic accuracy $\alpha_s (\alpha_s \ln x)^n$ 
can be written\cite{24} as follows:
\begin{equation}
  f_q^{(s)}(x,{\mathbf q}_T^2,\mu^2) = \int \limits_x^1 {dz \over z} \int d{\mathbf k}_T^2 
    {1\over {\mathbf \Delta}^2} {\alpha_s \over 2\pi} P_{qg}(z,{\mathbf k}_T^2,{\mathbf \Delta}^2) f_g(x/z,{\mathbf k}_T^2, \bar \mu^2),
\end{equation}

\noindent
where $z$ is the fraction of the gluon light cone momentum which is carried out by the quark,
and ${\mathbf \Delta} = {\mathbf q}_T - z{\mathbf k}_T$. 
The sea quark evolution is driven by the off-shell  gluon-to-quark splitting 
function $P_{qg}(z,{\mathbf k}_T^2,{\mathbf \Delta}^2)$\cite{25}:
\begin{equation}
  P_{qg}(z,{\mathbf k}_T^2,{\mathbf \Delta}^2) = T_R \left({\mathbf \Delta}^2\over {\mathbf \Delta}^2 + z(1-z)\,{\mathbf k}_T^2\right)^2 
    \left[(1 - z)^2 + z^2 + 4z^2(1 - z)^2 {{\mathbf k}_T^2\over {\mathbf \Delta}^2} \right],
\end{equation}

\noindent
with $T_R = 1/2$. The splitting function $P_{qg}(z,{\mathbf k}_T^2,{\mathbf \Delta}^2)$
has been obtained by generalizing to finite
transverse momenta, in the high-energy region, the two-particle irreducible kernel expansion\cite{47}.
Although evaluated off-shell, this splitting function is universal\cite{25}. It takes 
into account the small-$x$ enhanced transverse momentum dependence up to all orders in the strong coupling,
and reduces to the collinear splitting function at lowest order for ${\mathbf k}_T^2 \to 0$.
The scale $\bar \mu^2$ is defined\cite{24}
from the angular ordering condition which is natural from the point of view of the 
CCFM evolution: $\bar \mu^2 = {\mathbf \Delta}^2/(1-z)^2 + {\mathbf k}_T^2/(1-z)$.
To precise, in~(10) we have used A0 gluon\cite{46}.

Beside the CCFM-based approximation above, 
to determine the TMD quark densities in a proton
we have used also 
the Kimber-Martin-Ryskin (KMR) approach\cite{20,21}.
This approach is a formalism to construct the TMD parton distributions
from the known collinear ones. 
In this approximation, the TMD quark densities are given by\cite{20,21}
\begin{equation}
  \displaystyle f_q(x,{\mathbf q}_T^2,\mu^2) = T_q({\mathbf q}_T^2,\mu^2) {\alpha_s({\mathbf q}_T^2)\over 2\pi} \times \atop {
  \displaystyle \times \int\limits_x^1 dz \left[P_{qq}(z) {x\over z} q\left({x\over z},{\mathbf q}_T^2\right) \Theta\left(\varsigma - z\right) + P_{qg}(z) {x\over z} g\left({x\over z},{\mathbf q}_T^2\right) \right],} 
\end{equation}

\noindent
where $P_{ab}(z)$ are the unregulated leading-order DGLAP splitting 
functions. The theta function in~(12) implies the angular-ordering constraint $\varsigma = \mu/(\mu + |{\mathbf q}_T|)$ 
specifically to the last evolution step to regulate the soft gluon
singularities.
The Sudakov form factor $T_q({\mathbf q}_T^2,\mu^2)$ enable 
us to include logarithmic loop corrections
to the calculated cross sections. 
In the region of small ${\mathbf q}_T^2 < \mu_0^2$,
where $\mu_0^2 \sim 1$~GeV$^2$ is the minimum scale for which the DGLAP 
evolution of the initial parton densities is valid,
the TMD quark distributions are defined from the 
normalisation condition:
\begin{equation}
  f_q(x,{\mathbf q}_T^2,\mu^2)|_{{\mathbf q}_T^2 < \mu_0^2} = xq(x,\mu_0^2) T_q(\mu_0^2,\mu^2).
\end{equation}

\noindent
For the numerical calculations we have used the leading-order 
MSTW'2008 parton densities\cite{48}. 

The calculated TMD up, down and light sea quark densities 
are shown in Fig.~1 as a function of ${\mathbf q}_T^2$ for 
different values of $x$ at $\mu^2 = m_Z^2$.
Even with very different approaches,
the TMD quark densities are rather similar at large ${\mathbf q}_T^2$.
The influence of starting distributions and/or initial conditions is concentrated
at small values of ${\mathbf q}_T^2$.
It was pointed out\cite{18} that the small ${\mathbf q}_T^2$ region provides 
information on the non-perturbative part of the TMD parton density functions.
The difference between the CCFM-based and KMR approaches are visible  clearly
in the sea quark distributions which are driven mainly by the gluon densities.

\section{Numerical results} \indent

We now are in a position to present our numerical results. 
After we fixed the TMD quark 
densities, the cross section~(6) depends on the renormalization and 
factorization scales $\mu_R$ and $\mu_F$. Numerically, we set them to be 
equal to $\mu_R = \mu_F = \xi M$, where $M$ is the invariant mass of produced lepton pair. 
To estimate the scale uncertainties of our calculations we vary the parameter
$\xi$ between $1/2$ and $2$ about the default value $\xi = 1$.
Following to\cite{49}, we set $m_Z = 91.1876$~GeV, $\Gamma_Z = 2.4952$~GeV, 
$\sin^2 \theta_W = 0.23122$ and use the LO formula for the strong coupling constant
$\alpha_s(\mu^2)$ with $n_f= 4$ active quark flavors at
$\Lambda_{\rm QCD} = 200$~MeV, so that $\alpha_s (m_Z^2) = 0.1232$.
Since we investigate a wide region of $M$, we use the running QED coupling
constant $\alpha(\mu^2)$. To take into account the non-logarithmic loop corrections to the 
quark-antiquark annihilation cross section we apply the effective $K$-factor,
as it was done in\cite{21,37}:
\begin{equation}
  K = \exp \left[ C_F {\alpha_s(\mu^2)\over 2\pi} \pi^2 \right],
\end{equation}

\noindent
where color factor $C_F = 4/3$. A particular scale choice
$\mu^2 = {\mathbf p}_T^{4/3} M^{2/3}$ 
(with ${\mathbf p}_T$ being the transverse momentum of
produced lepton pair) has been proposed\cite{21,50}
to eliminate sub-leading logarithmic terms. We choose this scale 
to evaluate the strong coupling constant in~(14) only.
Everywhere the multidimensional integration have been performed 
by the means of Monte Carlo technique, using the routine \textsc{vegas}\cite{51}. 
The corresponding C++ code is available from the authors on request\footnote{lipatov@theory.sinp.msu.ru}.

Experimental data for the Drell-Yan production 
at the LHC come from the CMS\cite{28,29,30}, ATLAS\cite{31,32,33,34} and LHCb\cite{35} collaborations.
The CMS collaboration has reported the normalized dilepton invariant mass distribution 
measured at $15 < M < 600$~GeV\cite{28}. The normalized dilepton rapidity $y$ and 
transverse momentum distributions have been measured in the 
$Z$ boson mass region $60 < M <120$~GeV\cite{29}. 
The ATLAS and LHCb collaborations have presented
the differential cross sections
in the central ($|y| < 3.5$) and forward ($2 < y < 4.5$) dilepton rapidity regions
at $66 < M < 116$~GeV\cite{31,32,33,35}.
Very recently the ATLAS collaboration has measured the 
Drell-Yan differential cross-section as a function of dilepton 
invariant mass in the range $116 < M < 1500$~GeV\cite{34}.
Note that special cuts on the pseudorapidities and transverse momenta
of produced leptons have been applied in all 
these measurements, see\cite{28,29,30,31,32,33,34,35} for the detailed information.
Of course, we impose these cuts in the same manner as it was done
in the experimental analyses.

The results of our calculations are presented in Figs. 2 --- 6 in comparison 
with the LHC data. 
The differential cross sections as a function of dilepton invariant 
mass and rapidity are shown in Figs.~2 and~3. We find that these distributions are
described reasonably well by the CCFM-based calculations.
However, the KMR predictions significantly (by a factor of about 2) underestimate the LHC data,
mainly due to different behaviour of corresponding TMD sea quark densities at low transverse 
momenta (see Fig.~1).
We observe that the shape of dilepton invariant mass distributions is not very sensitive
to the TMD quark densities. This is in a contrast
with the distributions on the dilepton rapidity, where the CCFM and KMR 
predictions differ from each other (Fig.~3). 
The sensitivity of predicted cross sections to the TMD quark densities 
is also clearly visible in the transverse momentum distributions of
produced lepton pair, or in the distributions of the $\phi_\eta^*$ variable, as it
is shown in Figs.~4 and~5. The $\phi_\eta^*$ variable is defined as\cite{52,53,54}:
\begin{equation}
  \phi_\eta^* = \tan\left({\phi_{\rm acop} \over 2}\right) \left[ \cosh \left( {\Delta \eta \over 2} \right) \right]^{-1},
\end{equation}

\noindent
where $\phi_{\rm acop} = \pi - |\Delta \phi|$ is the acoplanarity angle, and 
$\Delta \eta$ and $\Delta \phi$ are the differences in pseudorapidity
and azimuthal angles between the leptons, respectively. The variable $\phi_\eta^*$ is correlated 
to the quantity
$|{\mathbf p}_T|/M$ and therefore probes the same physics as the 
dilepton transverse momentum\cite{55,56,57}. Note that both these observables are singular at leading order
in the collinear QCD approximation due to back-to-back kinematic,
whereas in the $k_T$-factorization approach the finite transverse 
momentum of lepton pair is generated already in simple quark-antiquark annihilation~(1).
One can see that none of the TMD quark densities under consideration describe well
the transverse momentum and $\phi_\eta^*$ distributions,
although their shapes are better reproduced by the KMR predictions.
The CCFM-based calculations overestimate the CMS\cite{29} and ATLAS\cite{32} data
(taken in a central dilepton rapidity region) at low transverse momenta and underestimate 
them if dilepton transverse momentum increases.
We note, however, that description of these observables is improved if 
higher order contributions (which, in particular, affect on the shape of 
tranverse momentum distributions) are taken into account\cite{37}.
In a forward dilepton rapidity region, the CCFM predictions agree well with the LHCb data (see Fig.~5).

According to~(2) and~(3), the off-shell amplitude of the Drell-Yan production 
subprocess contains both the vector and axial-vector couplings of $Z$ boson to fermions.
The corresponding differential cross section~(6) can be described by the polar and azimuthal 
angles of produced leptons in their rest frame. When integrated over 
the azimuthal angle, it can be presented as follows:
\begin{equation}
  {d\sigma \over d\cos \theta^*} \sim {3\over 8}(1 + \cos^2 \theta^*) + A_{\rm FB}\cos \theta^*,
\end{equation}

\noindent 
where $\theta^*$ is the emission angle of produced lepton with respect to the quark 
momentum in the dilepton rest frame, and 
$A_{\rm FB}$ is the parameter of forward-backward asymmetry. It is
defined as
\begin{equation}
  A_{\rm FB} = {\sigma_{\rm F} - \sigma_{\rm B} \over \sigma_{\rm F} + \sigma_{\rm B}},
\end{equation}

\noindent
where $\sigma_{\rm F}$ and $\sigma_{\rm B}$ are the total cross sections 
for forward and backward events, i.e. events with positive or negative values of $\cos \theta^*$.
At the dilepton invariant masses near the $Z$ boson peak, the asymmetry $A_{\rm FB}$ is predicted to be small due
to the small value of the lepton vector coupling.
Above and below the $Z$ boson peak, $A_{\rm FB}$ shows a characteristic energy dependence 
governed by $\gamma^*$ --- $Z$ interference. 
Deviations from the Standard Model predictions for $A_{\rm FB}$ may indicate 
the existence of new particles beyond the Standard Model.
Recently the CMS collaboration has presented\cite{30} a first measurement of the 
asymmetry $A_{\rm FB}$.
Our predictions for $A_{\rm FB}$ as a function of the dilepton invariant mass and rapidity
are shown in Fig.~6 in comparison with the CMS data.
Following to experimental procedure\cite{30}, we use the Collins-Soper frame where
$\theta^*$ is defined to be the angle between the lepton momentum and the axis that bisects the
angle between the direction of one proton and the direction opposite to the other proton.
We find a good agreement of our predictions and the CMS data as was also obtained with collinear QCD
calculations\cite{30}. In contrast with the dilepton
transverse momentum and rapidity distributions,
there is practically no differences between the 
CCFM-based and KMR predictions for $A_{\rm FB}$. Note 
that the angular distributions in dilepton production at the Tevatron
have been investigated in\cite{37}.

To conclude, we have demonstrated that the studies of Drell-Yan production
(in particular, investigations of the dilepton transverse momentum and
rapidity distributions) impose a stringent constraints on the TMD quark densities in a proton.
Moreover, this process can be used as an important tool to determine 
the parameters of initial (starting) TMD parton distributions.
It is important for further investigations of small-$x$ physics at hadron colliders,
in particular, in the direction which concerns the non-linear effects originating 
from high parton densities at small $x$.

\section{Conclusion} \indent

We used the TMD quark densities in a proton to
investigate unpolarized Drell-Yan lepton pair production in 
$pp$ collisions at the LHC energies.
We investigated the case where the gluon-to-quark splitting occurs at the last evolution step
and calculated the TMD sea quark density as a convolution of the CCFM-evolved gluon 
distribution and TMD gluon-to-quark splitting function.
This function contains all single logarithmic small-$x$ corrections
to sea quark evolution for any order of perturbation theory.
We calculated  ${\cal O}(\alpha)$ production 
amplitude $q^* + \bar q^* \to Z/\gamma^* \to l^+ + l^-$ 
 within  the reggeized (off-shell) quark approach
which ensures the exact gauge invariance.
The higher-order ${\cal O}(\alpha\alpha_s)$
subprocesses $q^* + g^* \to Z/\gamma^* + q \to l^+ + l^- + q$ and
$q^* + \bar{q}^* \to Z/\gamma^* + g \to l^+ + l^- +g$ are present in
calculations as part of the evolution of TMD parton densities.
We have  analyzed the distributions on the dilepton invariant mass,
transverse momentum and rapidity as well as 
the specific angular correlations between the produced leptons 
as measured by the CMS, ATLAS and LHCb collaborations.
These measurements impose a stringent 
constraints on the TMD quark distributions in a proton. We obtain 
a reasonably good description of the experimental measurements with our approach.

\section*{Acknowledgements} \indent

We thank H.~Jung for careful
reading the manuscript and very useful remarks.
The authors are also grateful to F.~Hautmann and S.~Marzani for 
discussions and comments.
This research was supported by the FASI of Russian Federation
(grant NS-3042.2014.2), RFBR grant 13-02-01060 and the grant of the 
Ministry of education and sciences
of Russia (agreement 8412).
The authors are also grateful to DESY Directorate for the
support in the framework of Moscow---DESY project on Monte-Carlo implementation for
HERA---LHC.

\newpage

\begin{figure}
\begin{center}
\epsfig{figure=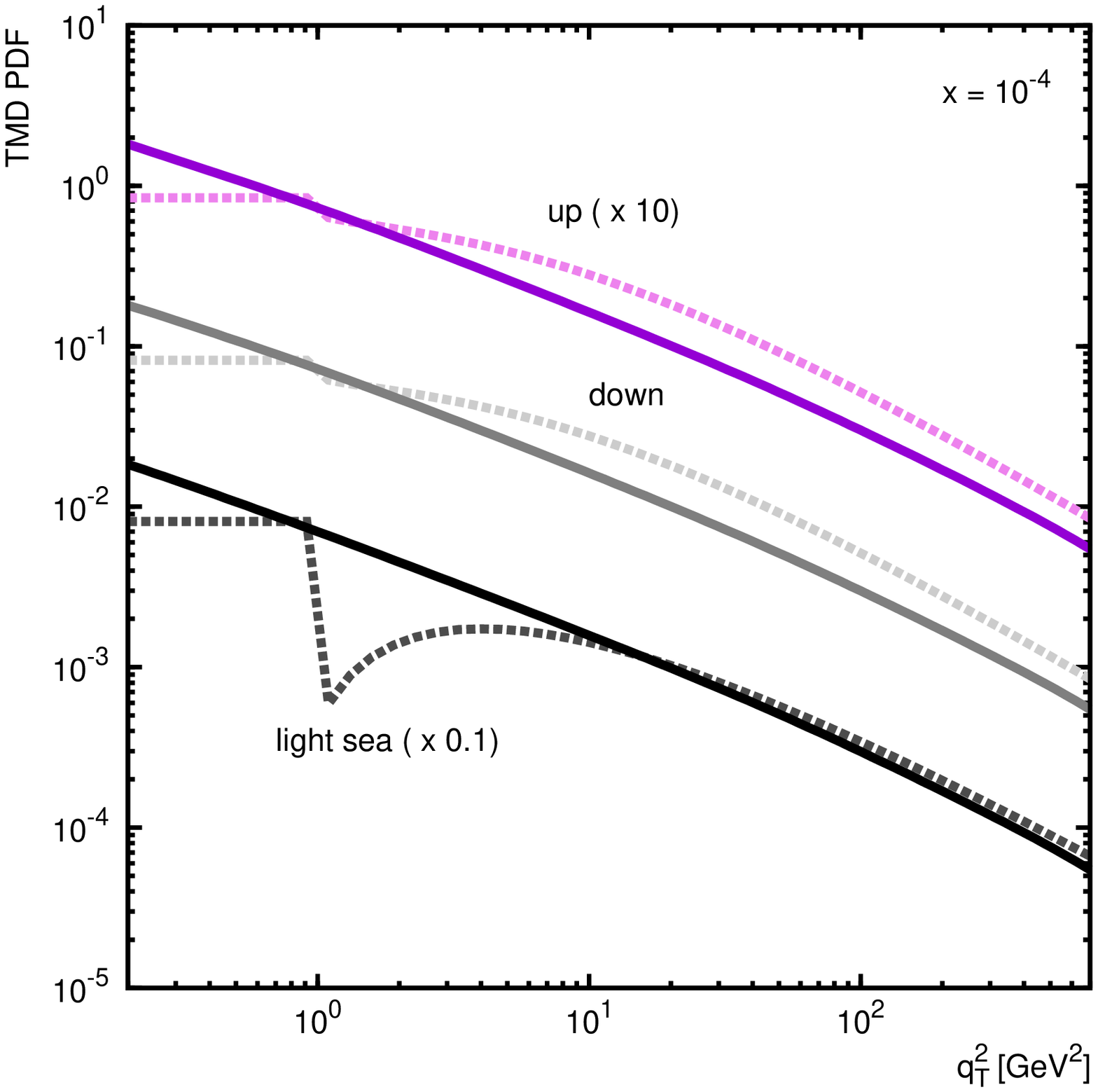, width = 5.5cm, angle = 270}
\epsfig{figure=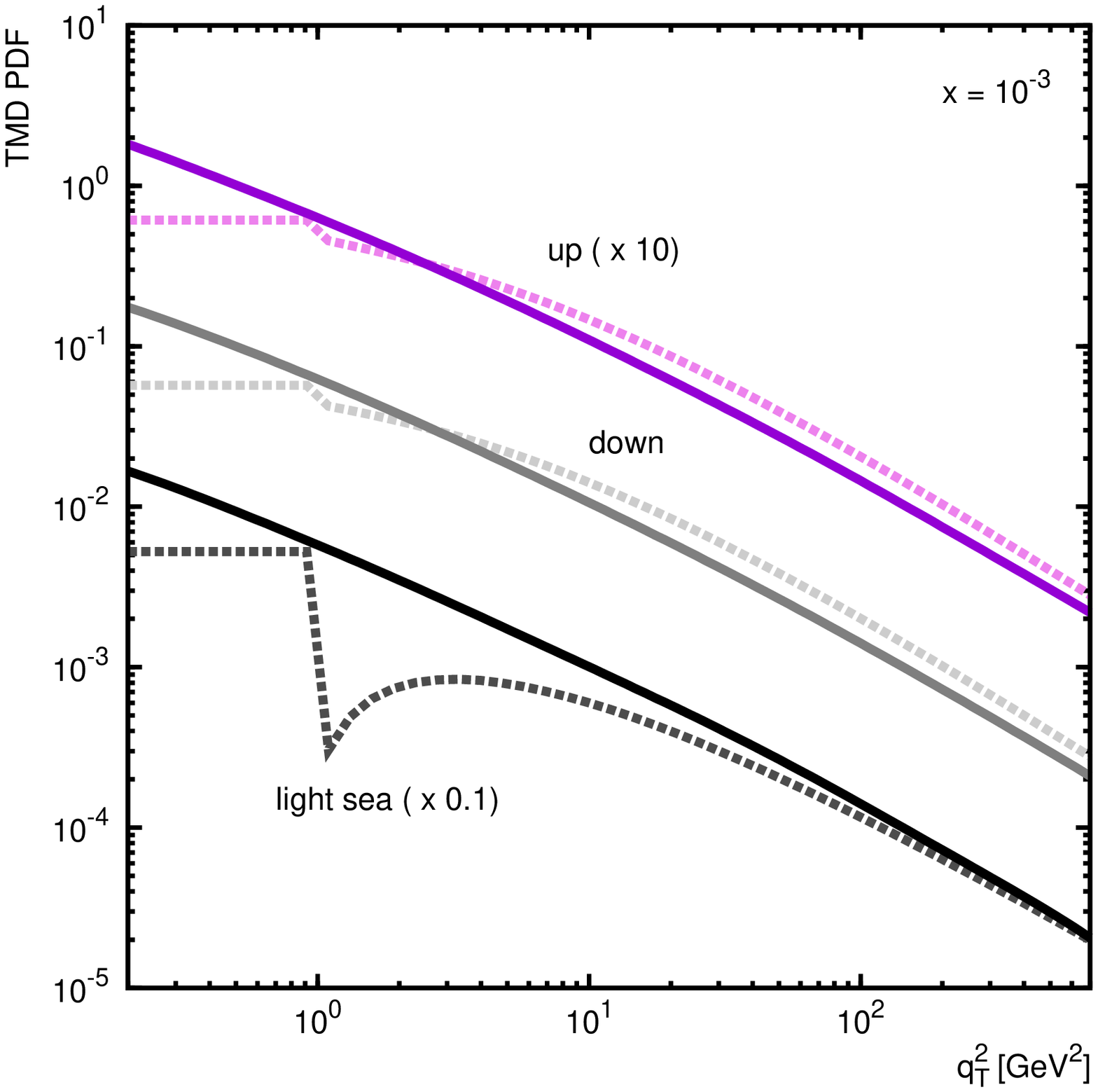, width = 5.5cm, angle = 270}
\epsfig{figure=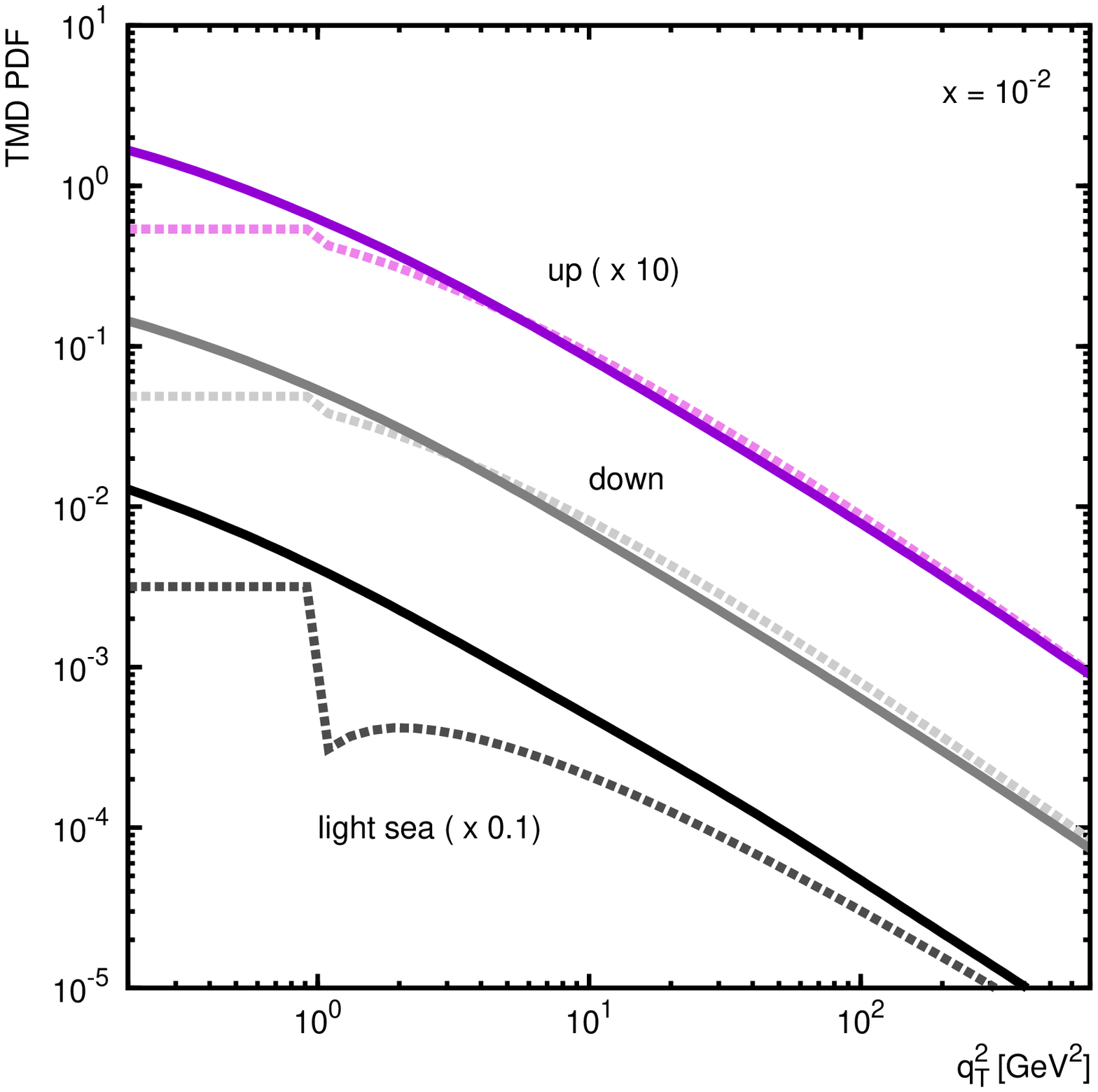, width = 5.5cm, angle = 270}
\epsfig{figure=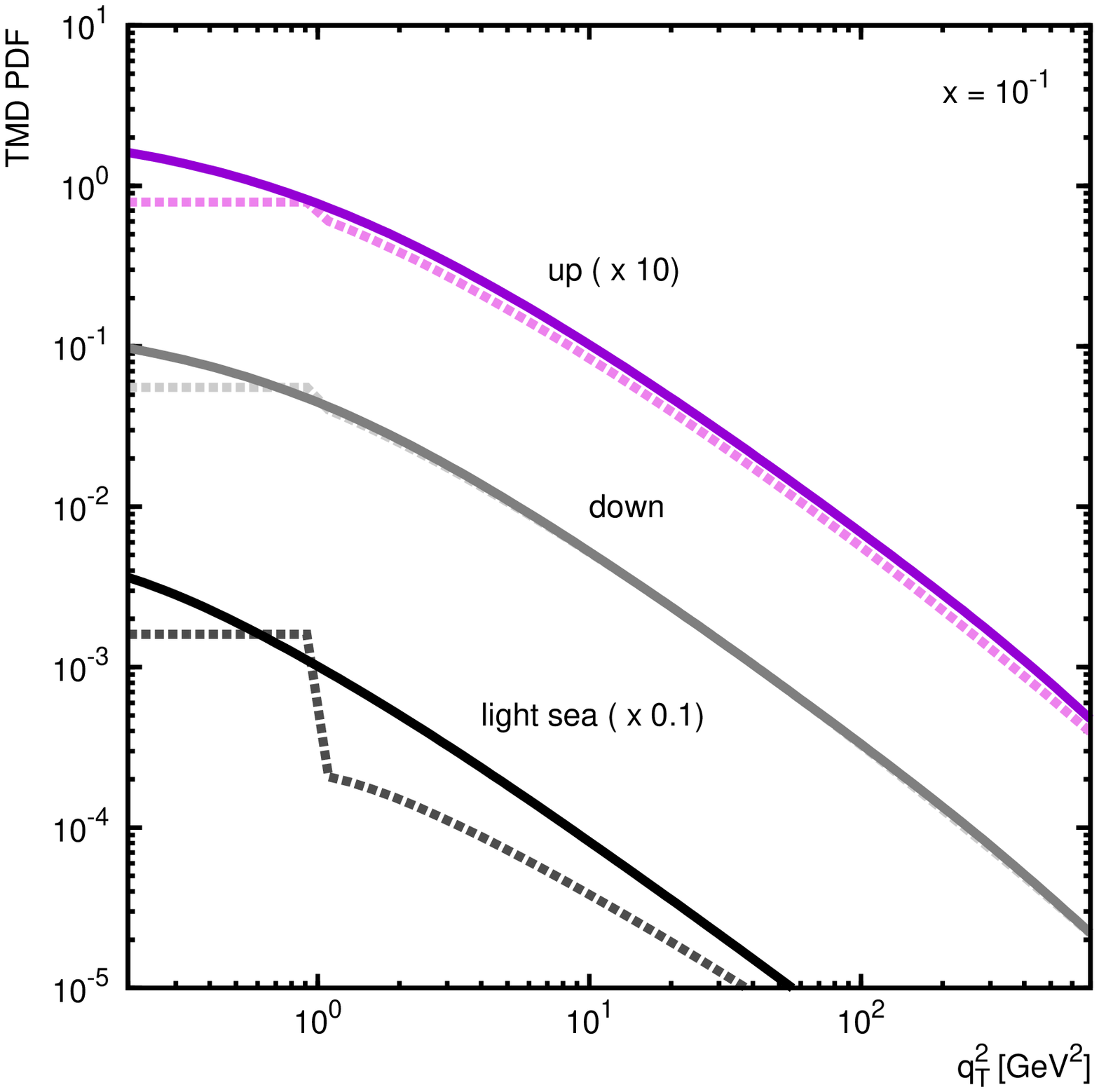, width = 5.5cm, angle = 270}
\caption{The TMD quark densities $f_q(x,{\mathbf q}_T^2,\mu^2)$ calculated 
as a function of quark transverse momentum ${\mathbf q}_T^2$ at several fixed $x$ values and $\mu^2 = m_Z^2$. 
The solid and dashed curves correspond 
to the CCFM-based and KMR quark densities, respectively.}
\label{fig1}
\end{center}
\end{figure}

\begin{figure}
\begin{center}
\epsfig{figure=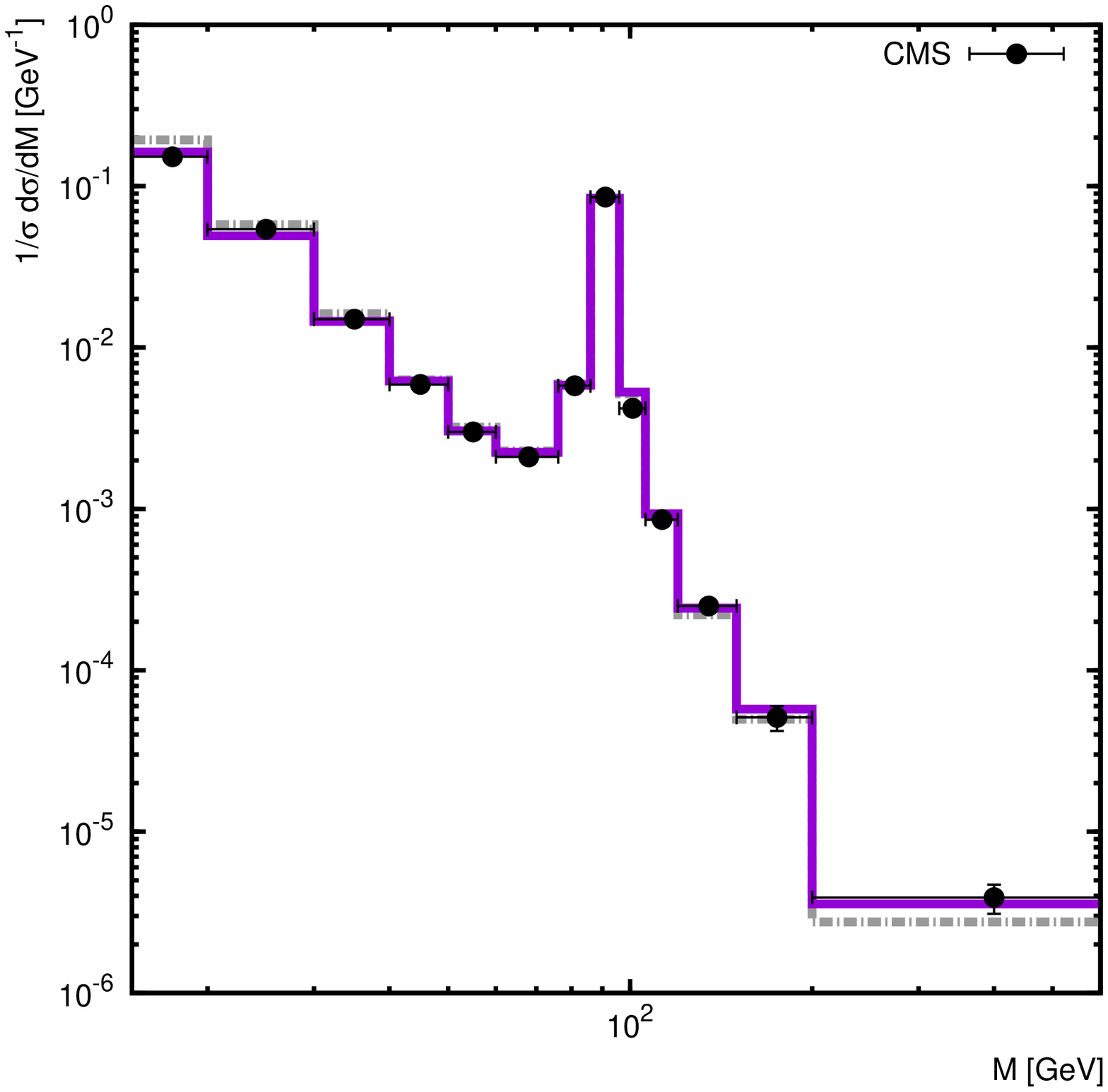, width = 5.5cm, angle = 270}
\epsfig{figure=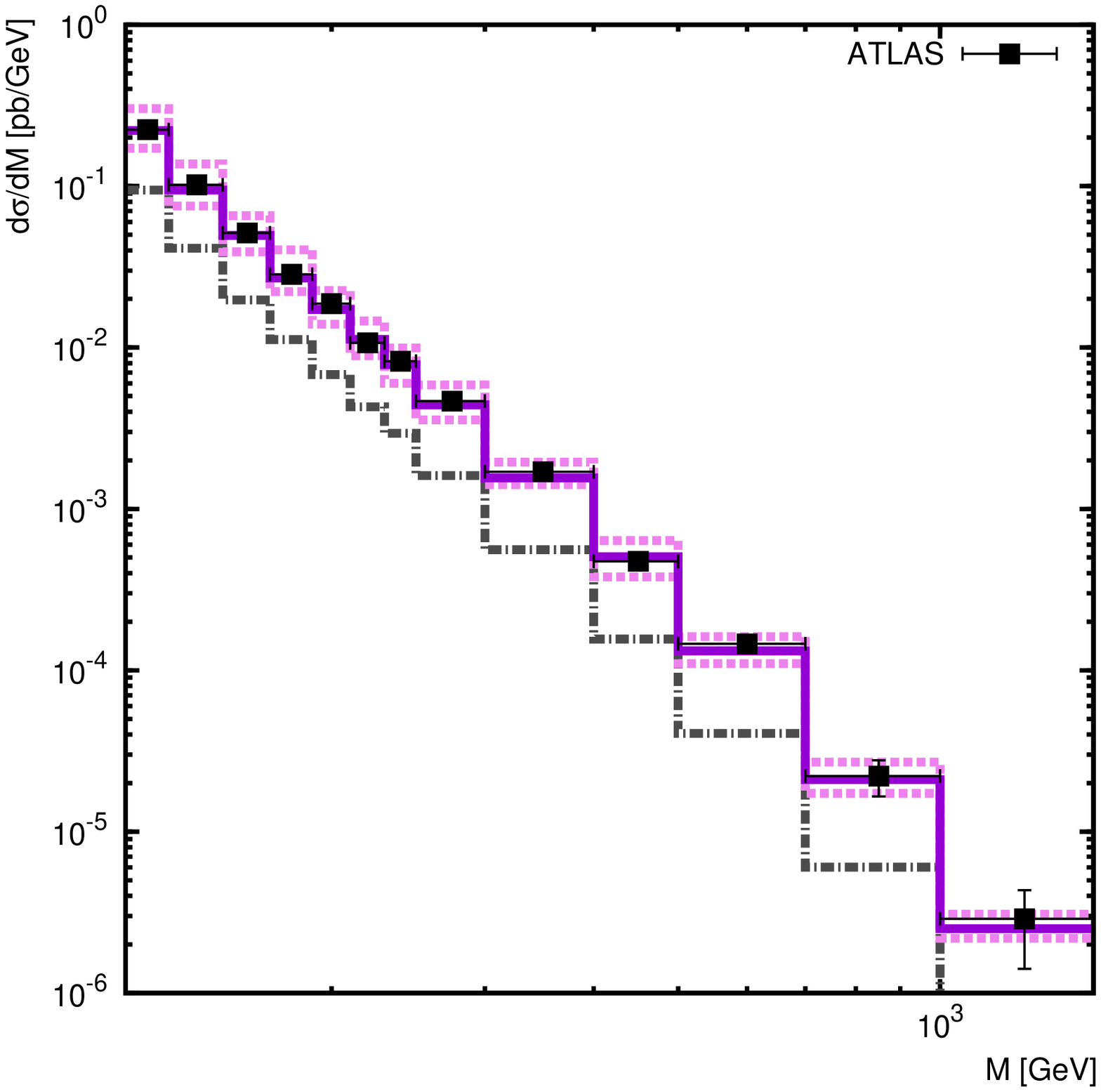, width = 5.5cm, angle = 270}
\caption{The differential cross sections of Drell-Yan lepton pair production in $pp$ collisions 
at the LHC as a function of dilepton invariant mass. The solid and dash-dotted
histograms correspond to the CCFM-based and KMR predictions,
respectively. The upper and lower dashed histograms correspond to the scale variations in the 
CCFM calculations, as it is described in the text. 
The experimental data are from CMS\cite{28} and ATLAS\cite{34}.}
\label{fig2}
\end{center}
\end{figure}

\begin{figure}
\begin{center}
\epsfig{figure=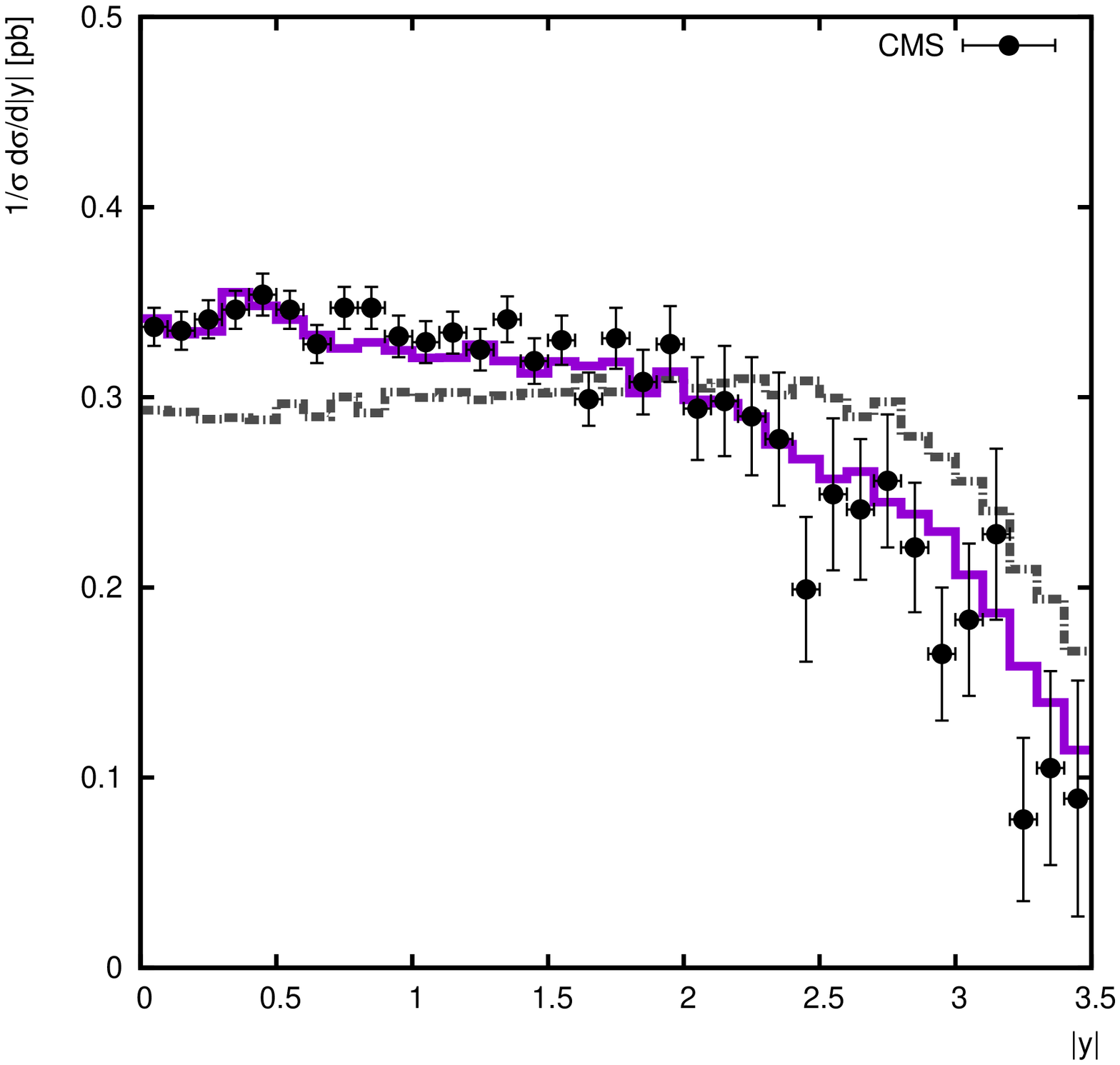, width = 5.5cm, angle = 270}
\epsfig{figure=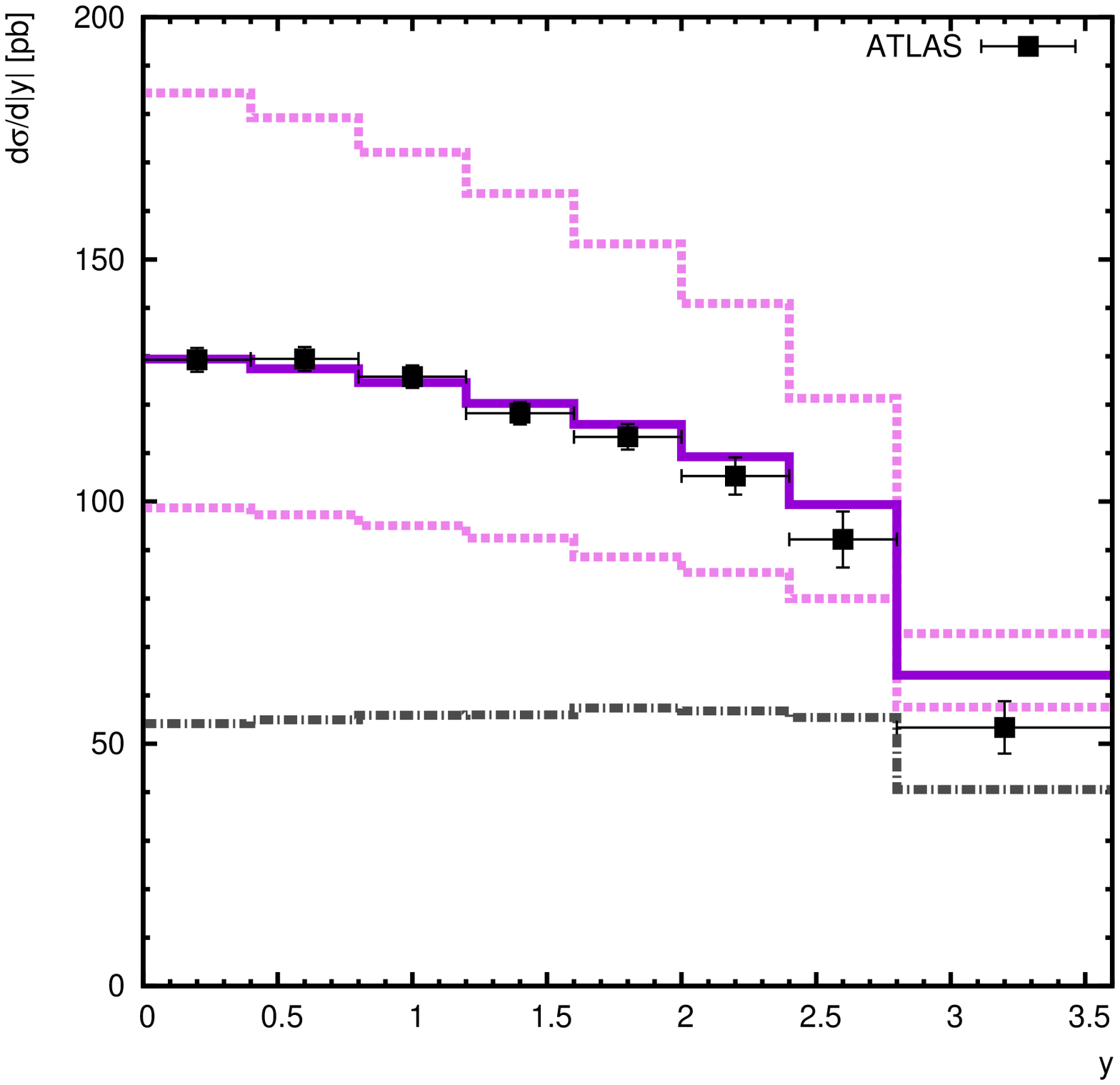, width = 5.5cm, angle = 270}
\epsfig{figure=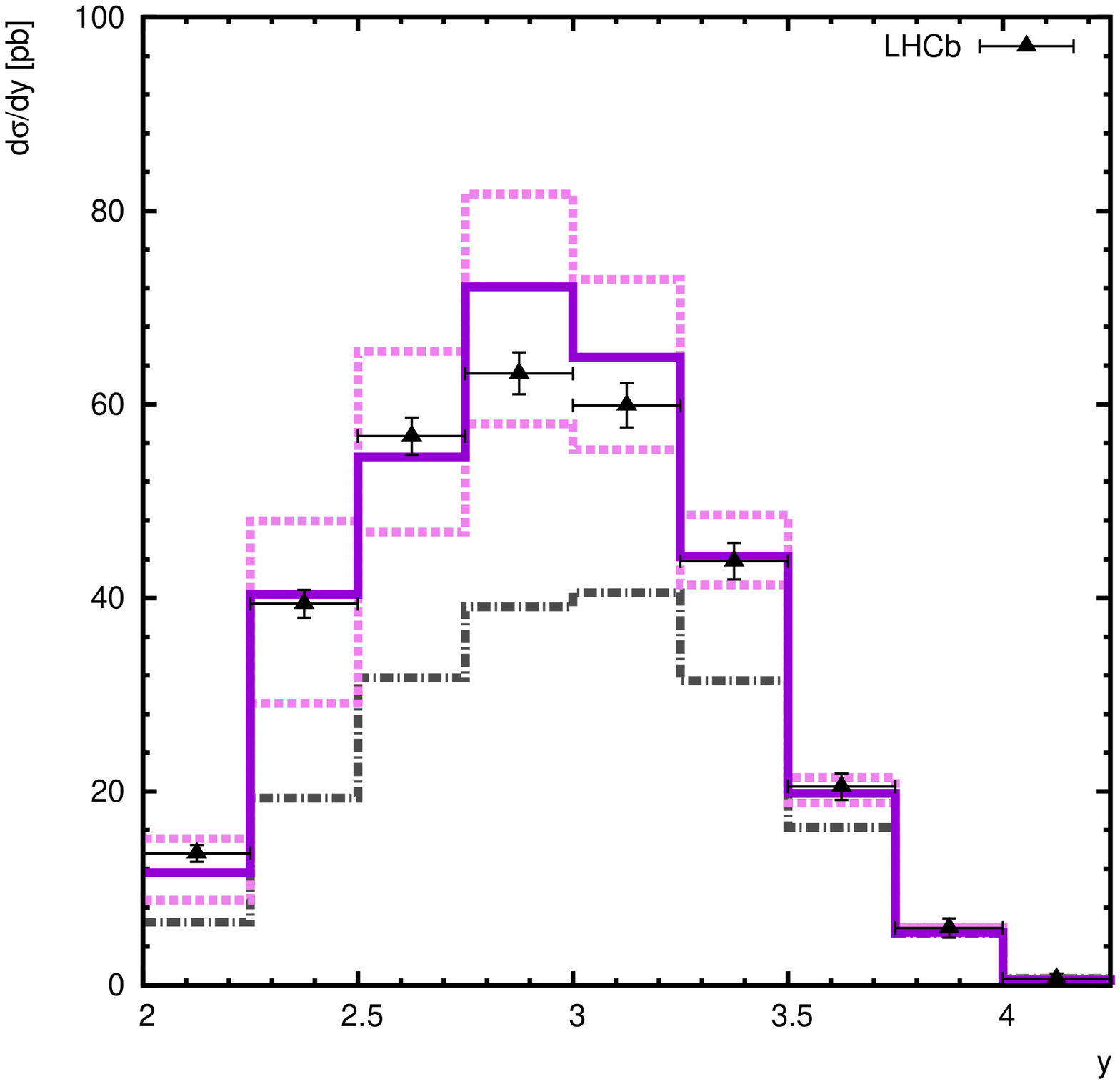, width = 5.5cm, angle = 270}
\caption{The differential cross sections of Drell-Yan lepton pair production in $pp$ collisions 
at the LHC as a function of dilepton rapidity $y$. Notation of all histograms is the same
as in Fig.~2. The experimental data are from CMS\cite{29}, ATLAS\cite{32} and LHCb\cite{35}.}
\label{fig3}
\end{center}
\end{figure}

\begin{figure}
\begin{center}
\epsfig{figure=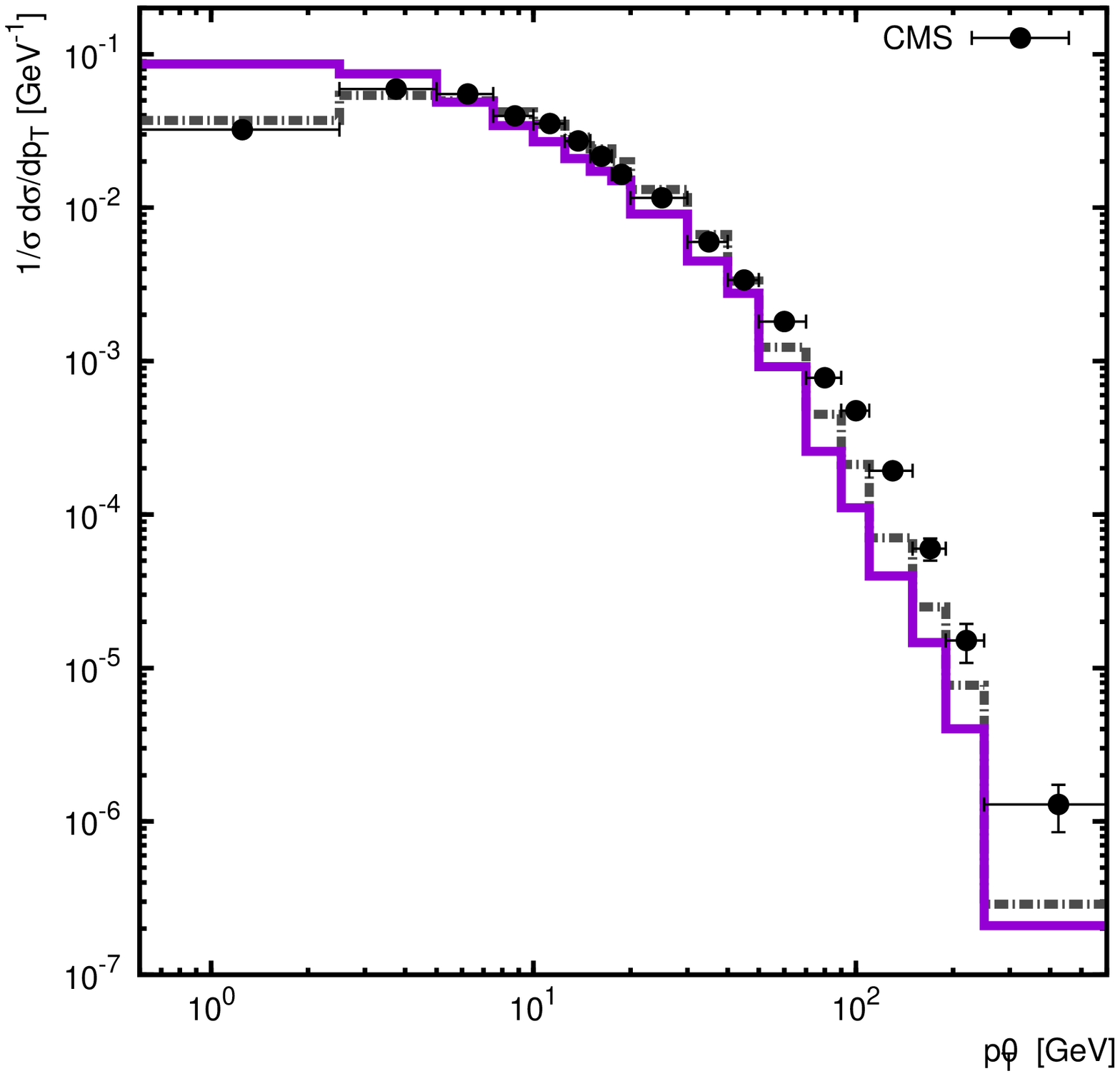, width = 5.5cm, angle = 270}
\epsfig{figure=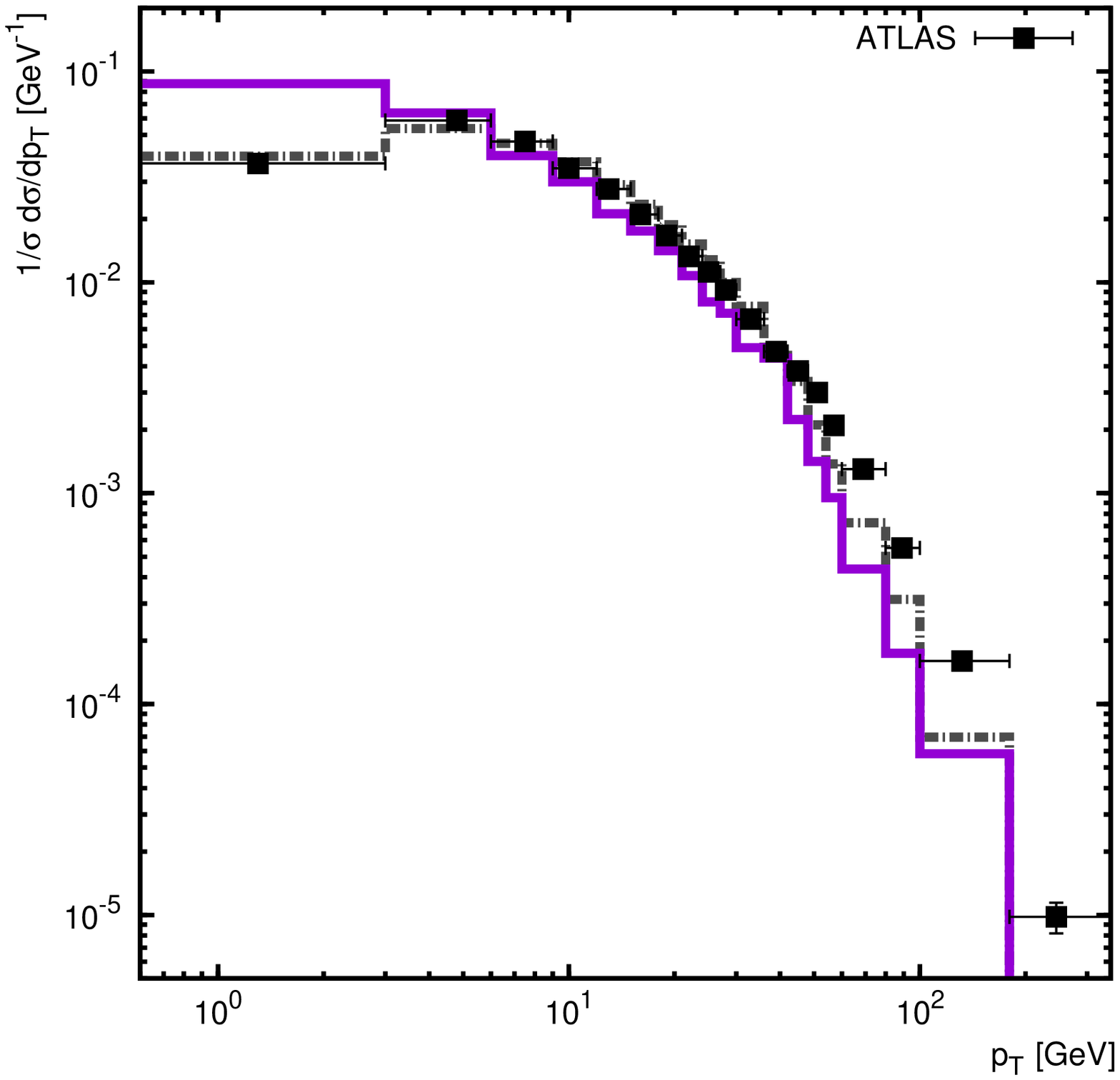, width = 5.5cm, angle = 270}
\caption{The differential cross sections of Drell-Yan lepton pair production in $pp$ collisions 
at the LHC as a function of dilepton transverse momentum. Notation of all histograms is the same
as in Fig.~2. The experimental data are from CMS\cite{29} and ATLAS\cite{31}.}
\label{fig4}
\end{center}
\end{figure}

\begin{figure}
\begin{center}
\epsfig{figure=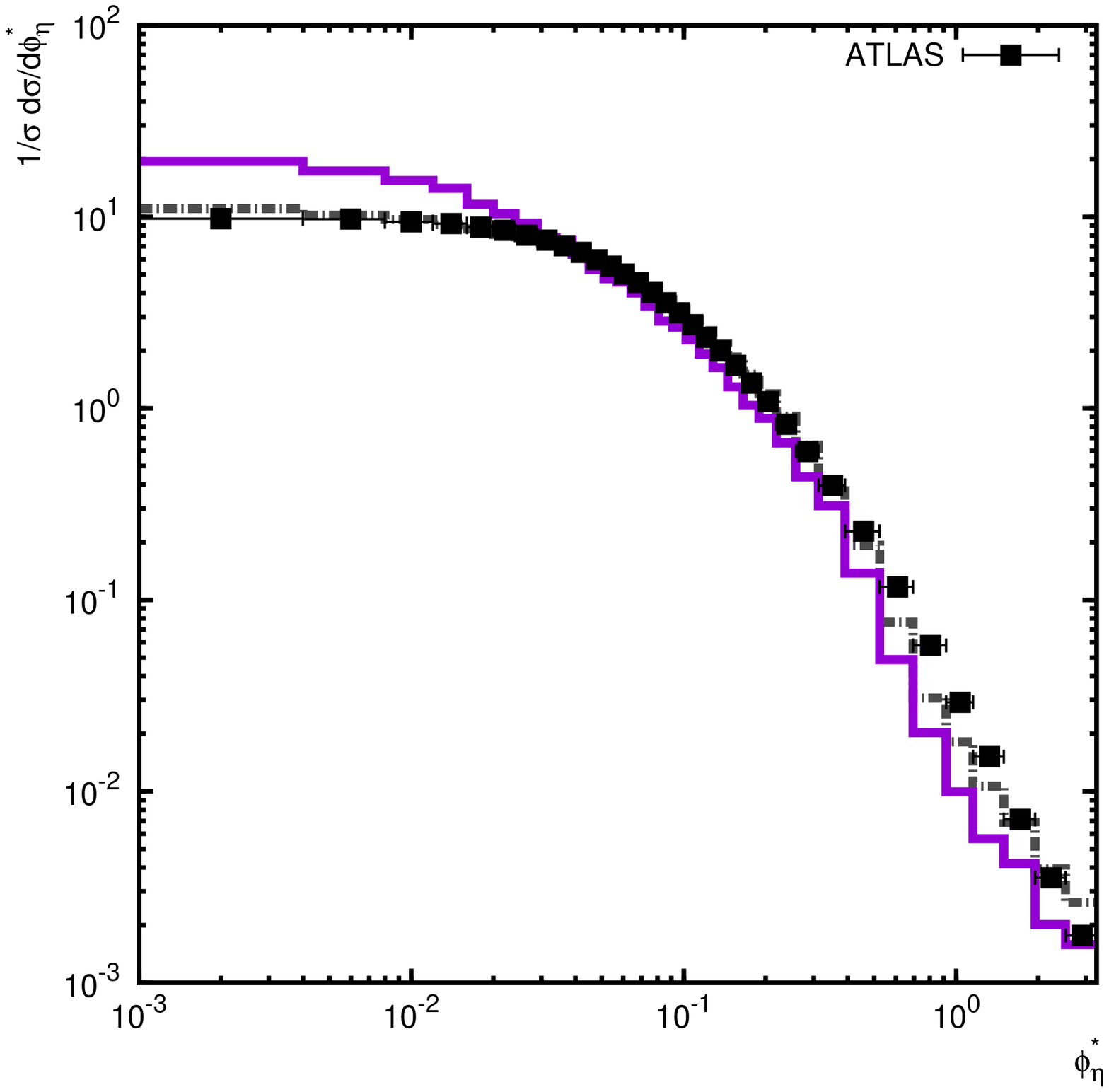, width = 5.5cm, angle = 270}
\epsfig{figure=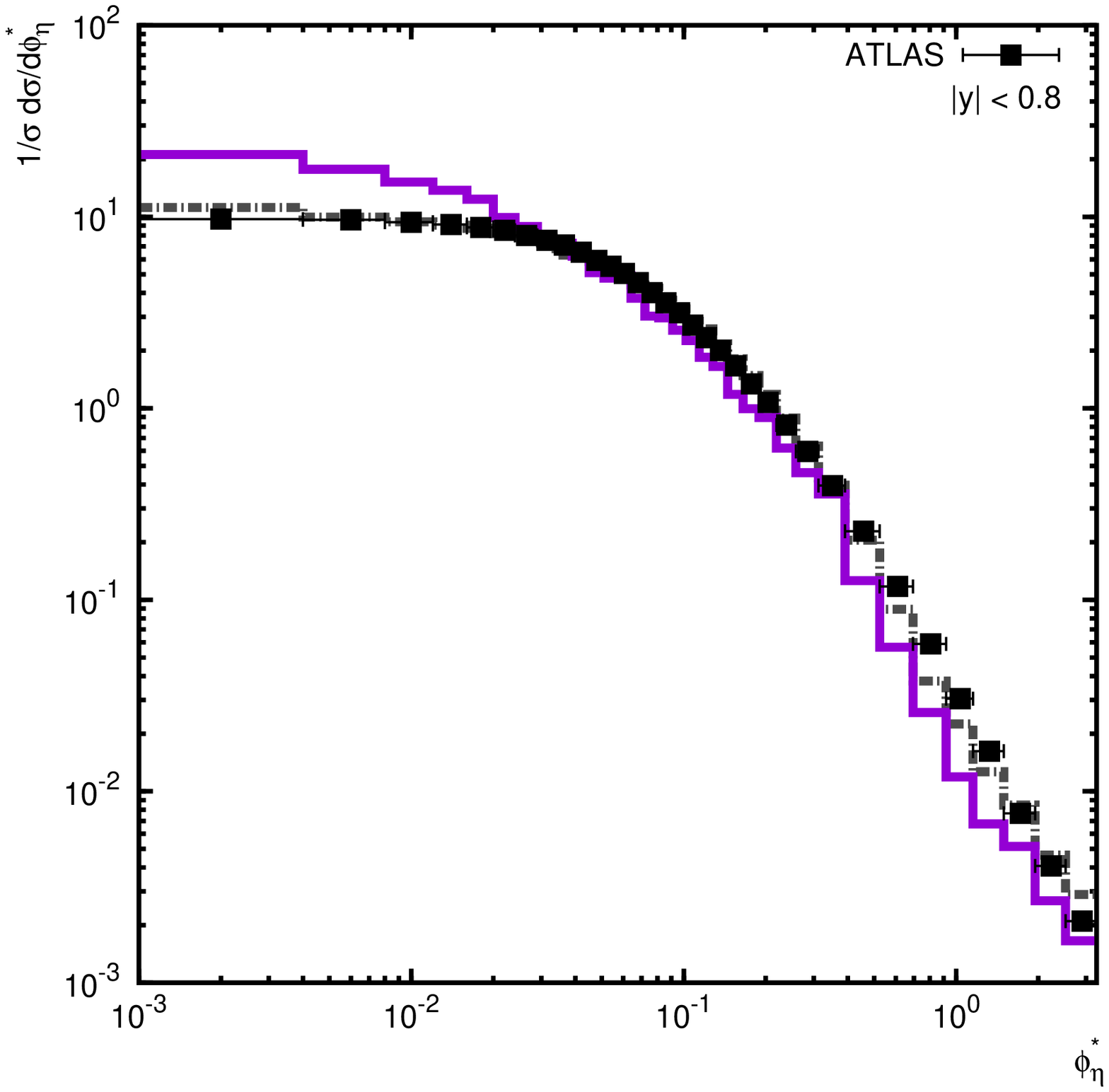, width = 5.5cm, angle = 270}
\epsfig{figure=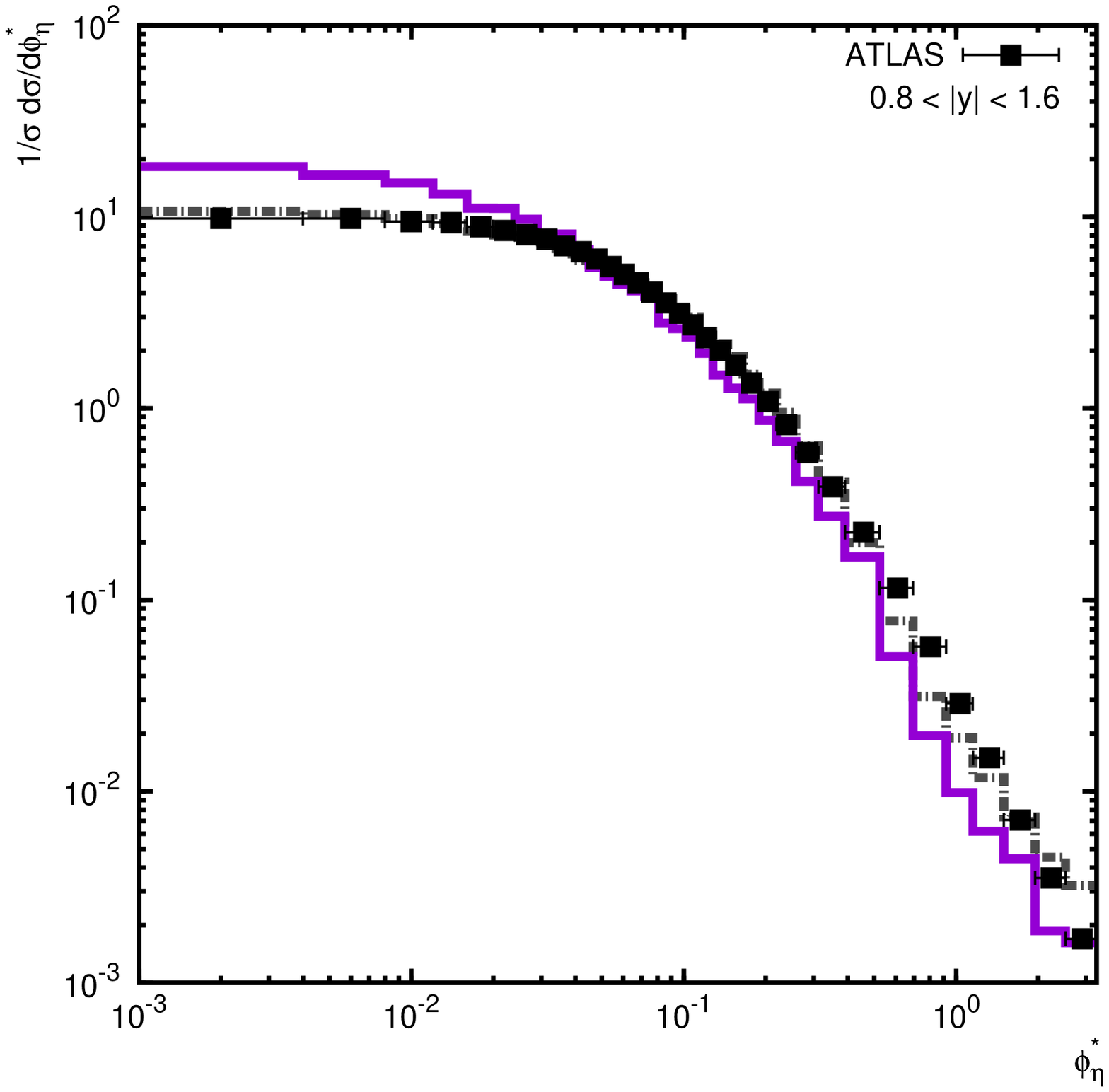, width = 5.5cm, angle = 270}
\epsfig{figure=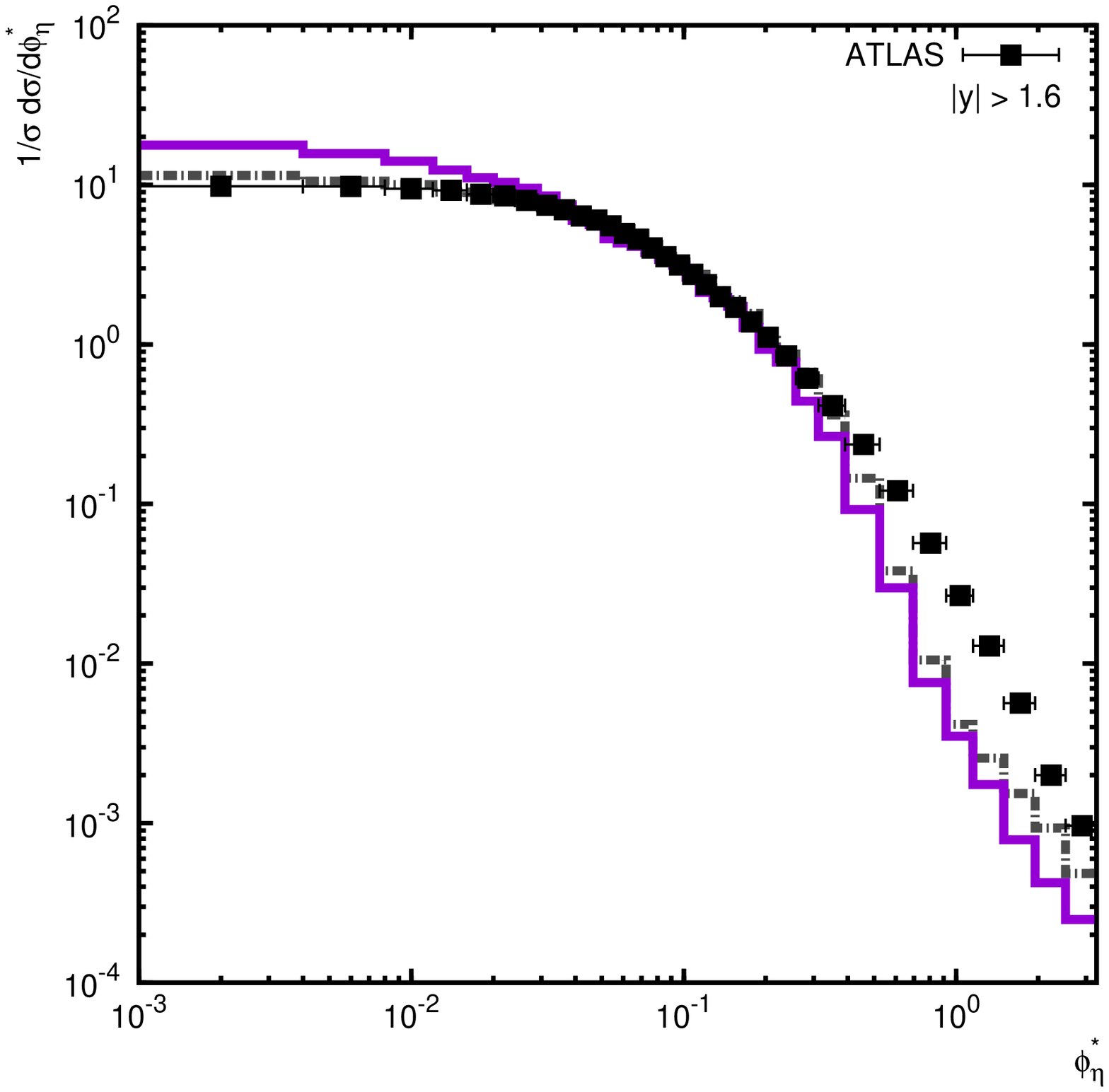, width = 5.5cm, angle = 270}
\epsfig{figure=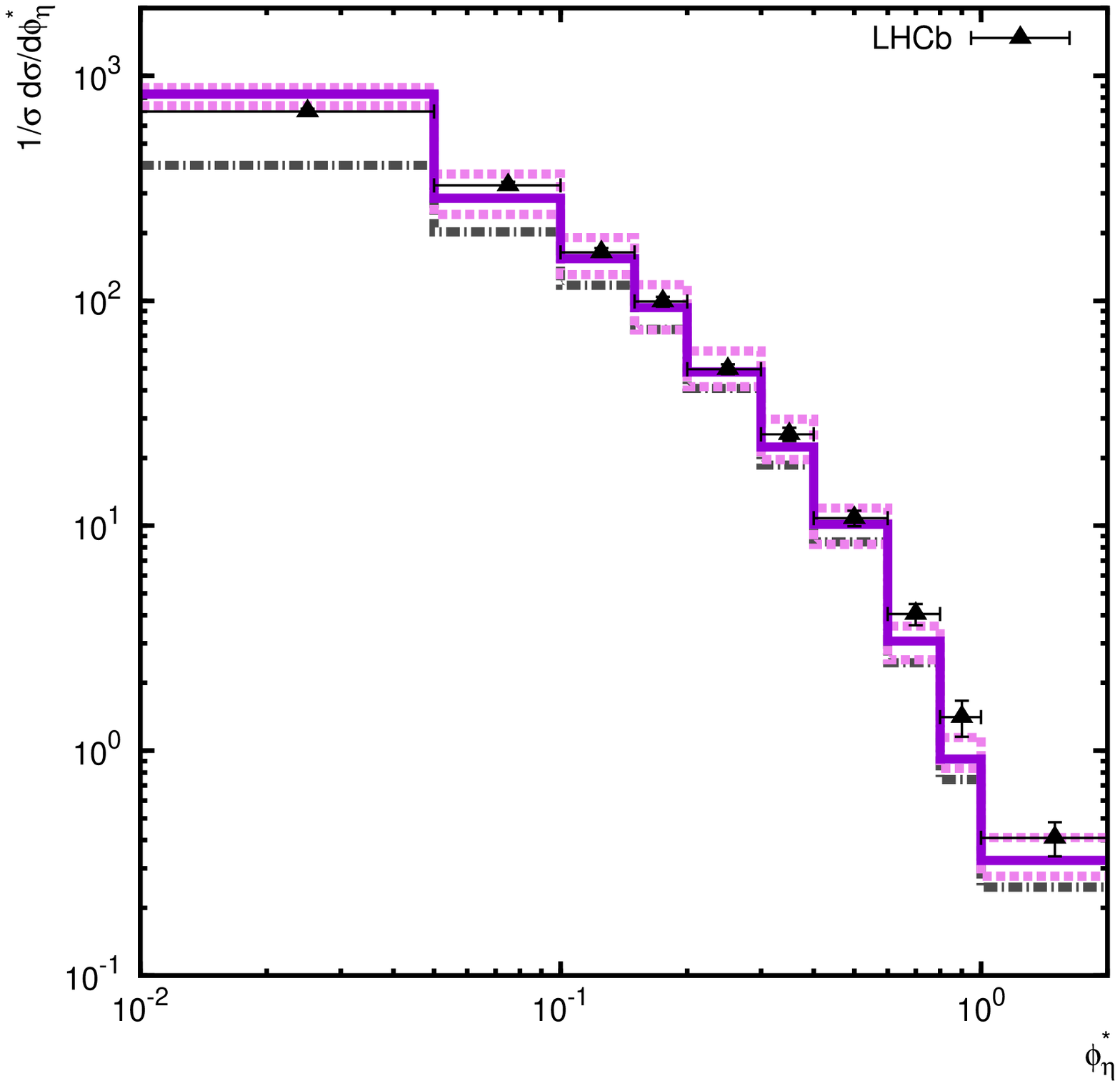, width = 5.5cm, angle = 270}
\caption{The differential cross sections of Drell-Yan lepton pair production in $pp$ collisions 
at the LHC as a function of $\phi_\eta^*$. Notation of all histograms is the same
as in Fig.~2. The experimental data are from ATLAS\cite{33} and LHCb\cite{35}.}
\label{fig5}
\end{center}
\end{figure}

\begin{figure}
\begin{center}
\epsfig{figure=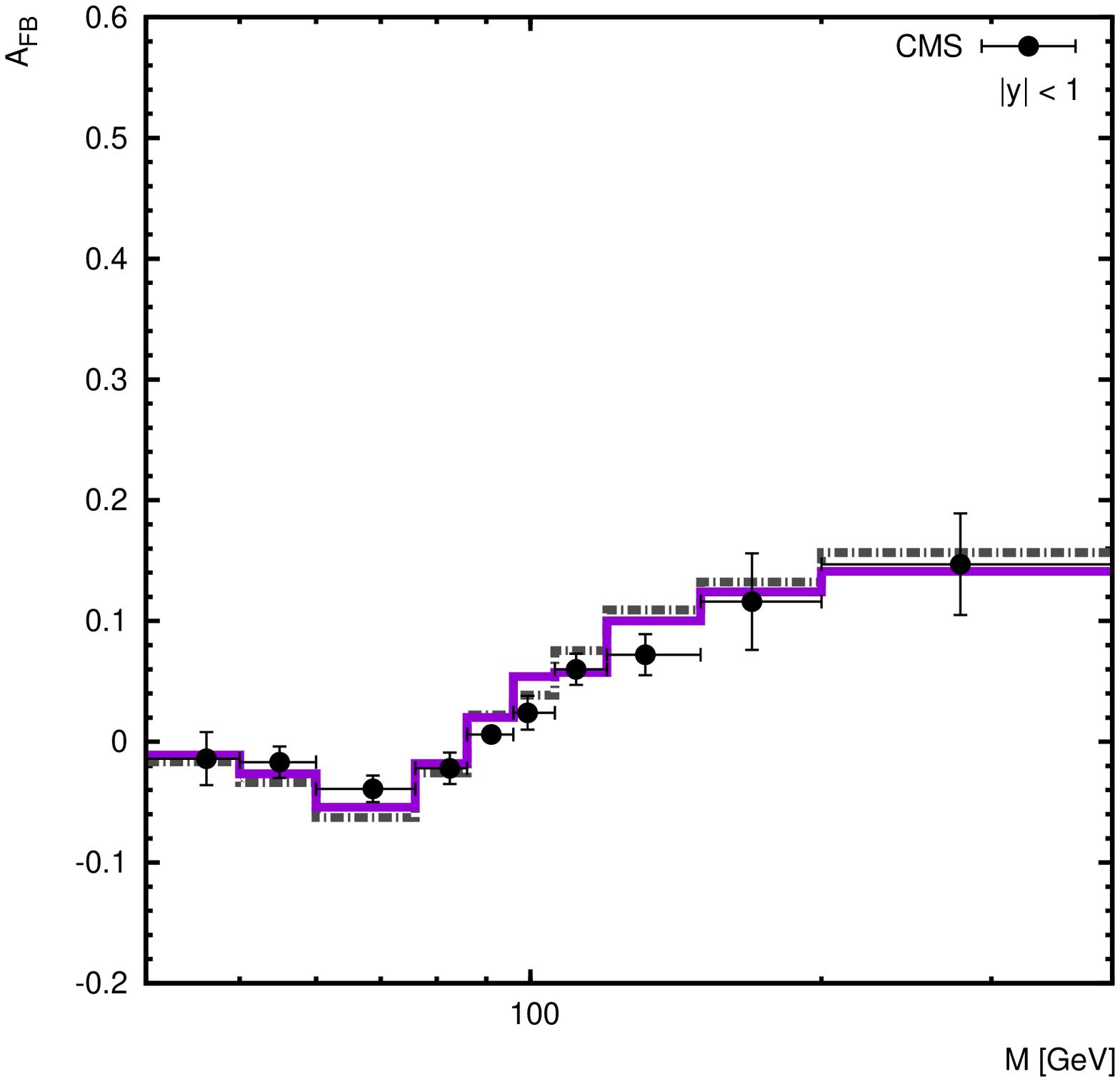, width = 5.5cm, angle = 270}
\epsfig{figure=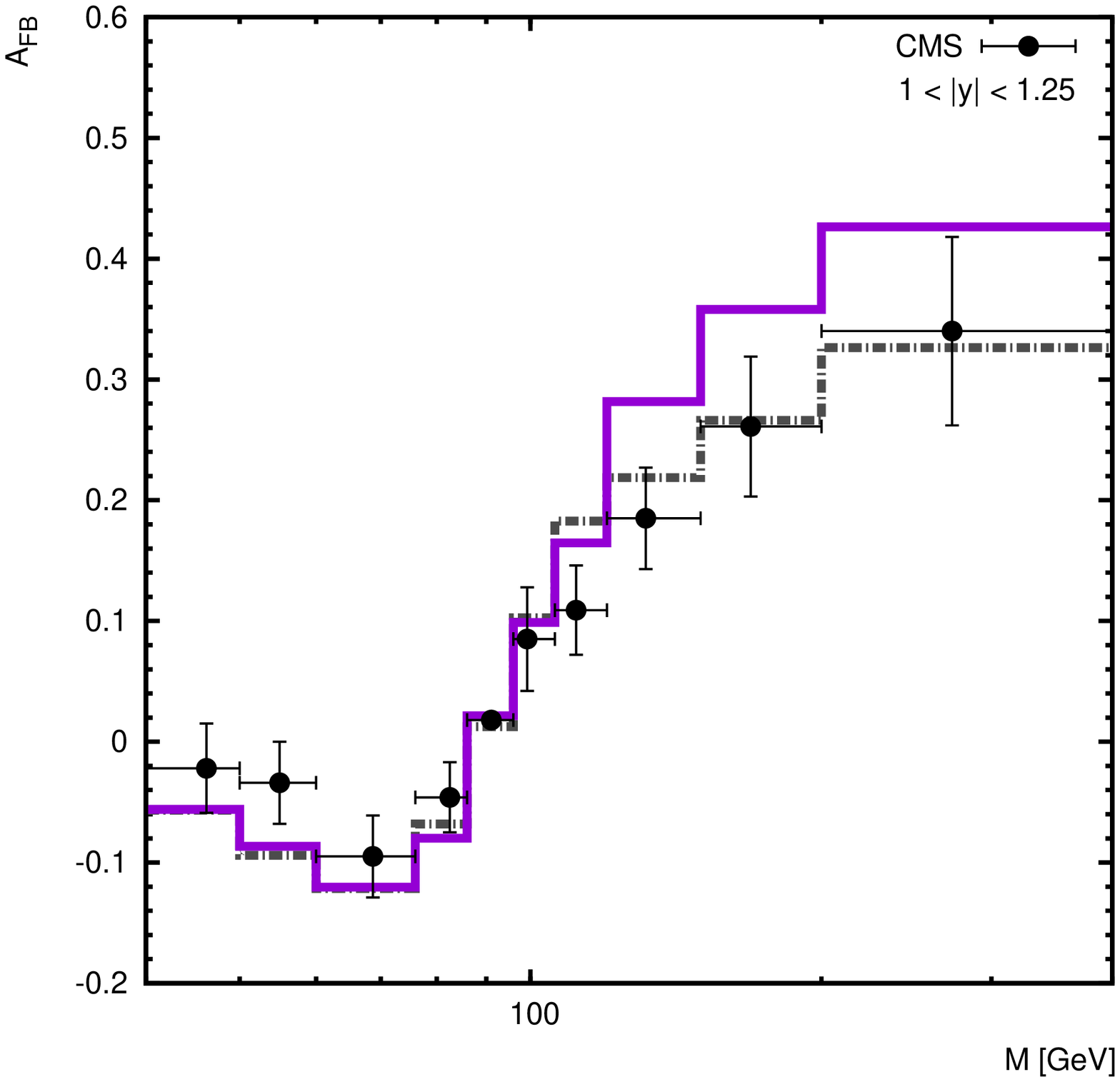, width = 5.5cm, angle = 270}
\epsfig{figure=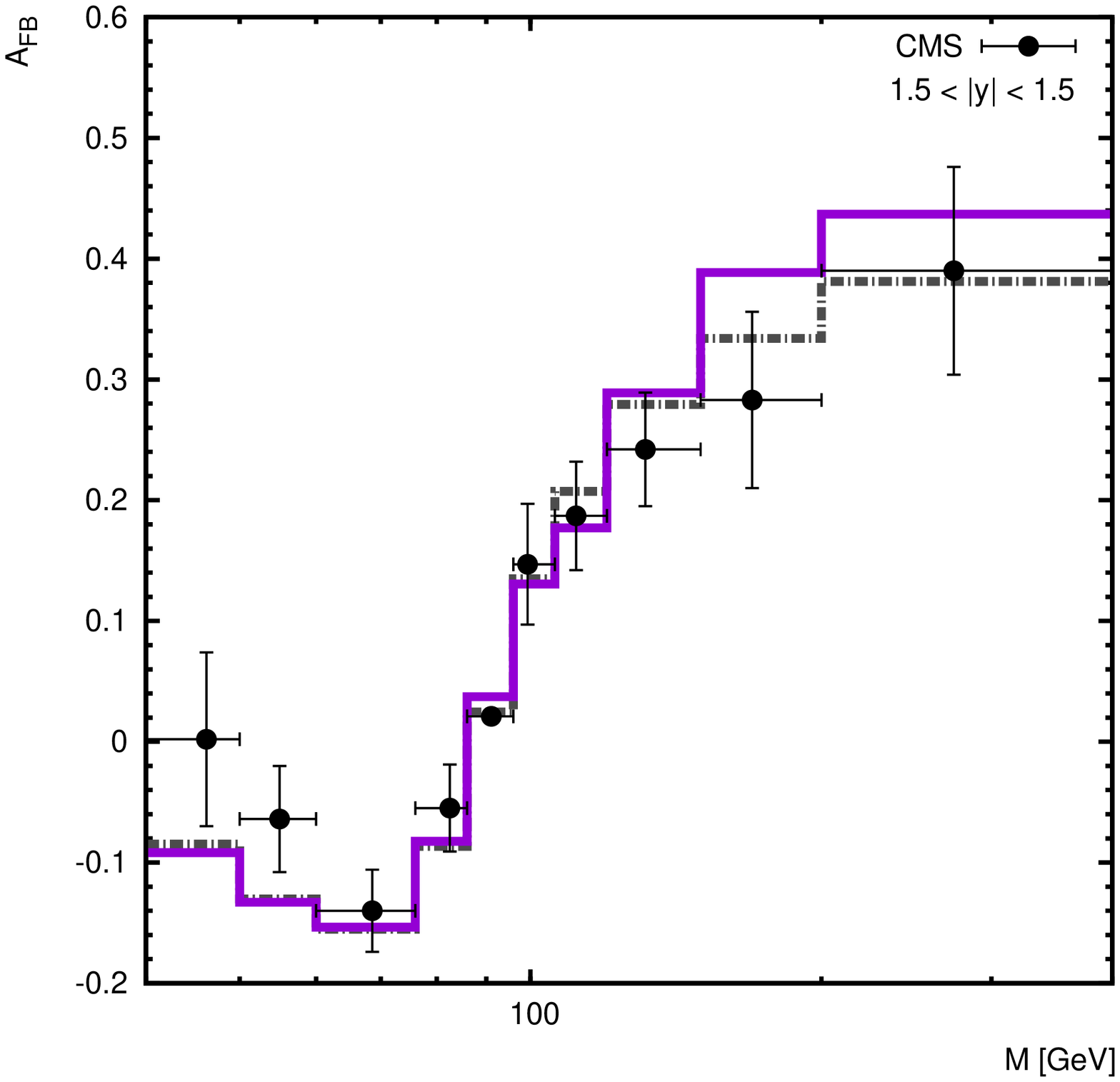, width = 5.5cm, angle = 270}
\epsfig{figure=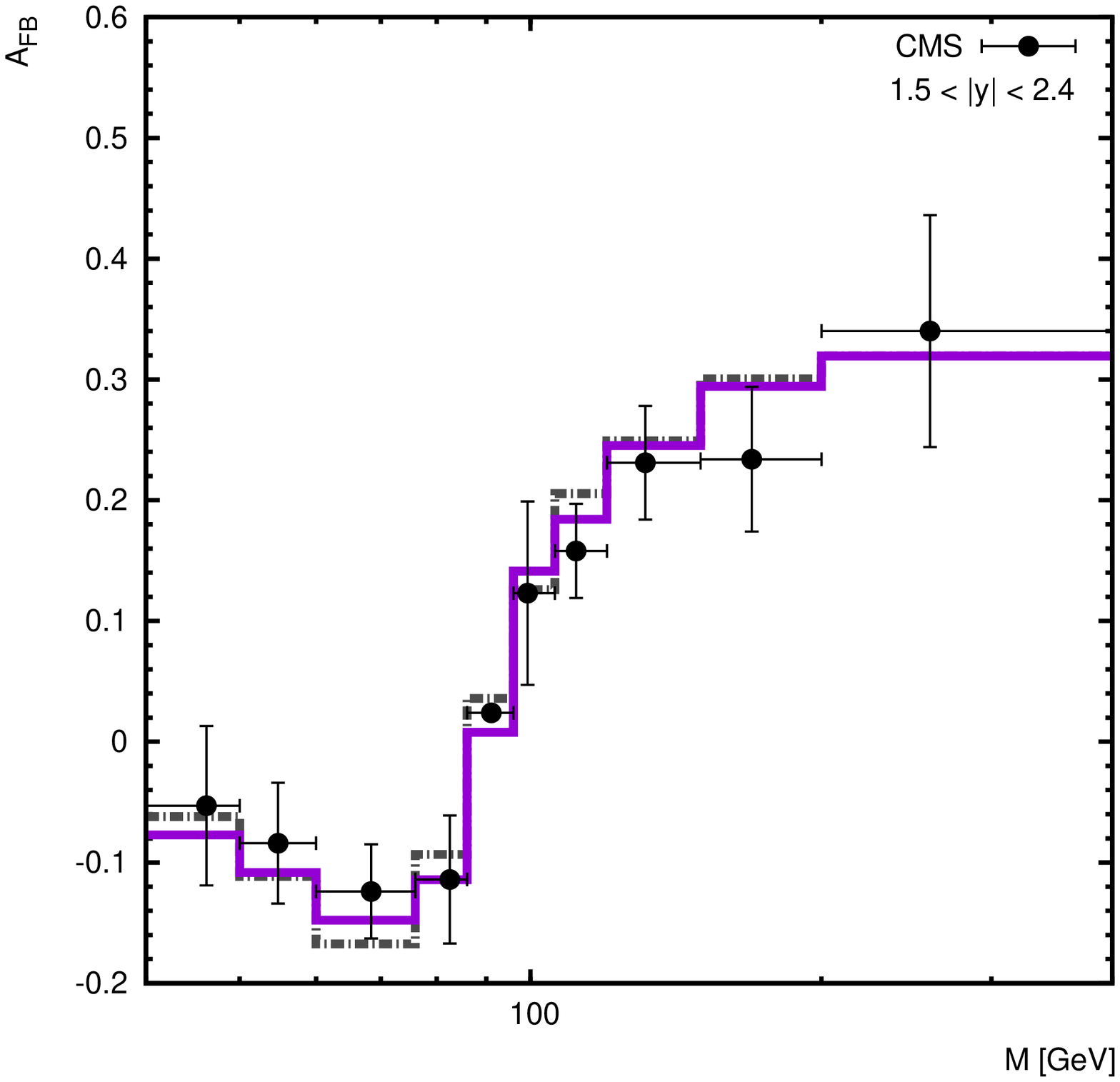, width = 5.5cm, angle = 270}
\caption{The forward-backward asymmetry $A_{\rm FB}$ calculated 
as a function of dilepton rapidity and invariant mass. 
Notation of all histograms is the same
as in Fig.~2. The experimental data are from CMS\cite{30}.}
\label{fig6}
\end{center}
\end{figure}

\end{document}